# Polarity-dependent electro-wetting or -dewetting on a conductive silicon substrate

Lele Zhou[a], Rong Zhang[a], Hung-Ta Wang[b], Yifan Liu[a]

[a]Division of Chemistry and Physical Biology, School of Physical Science and Technology, ShanghaiTech University, Shanghai, 201210, China

[b]School of Physical Science and Technology, ShanghaiTech University, 393 Middle Huaxia Road, Pudong, Shanghai 201210, China

**We demonstrate droplet manipulation using electric signals to induce the liquid to wet or dewet on a hydrophilic conductive substrate in air without adding layers. In this phenomenon, the contact angle changes more than 15° or −20° by using the ionic surfactant-mediated droplets, with only $\pm 3$ volts of the actuation voltage.**

Changes in the wettability of liquids under electric fields have long been studied[1]. A hydrophobic surface can become hydrophilic during droplet manipulation using the external electric field and vice versa[2]. Both electrowetting (EW) and electrowetting-on-dielectric (EWOD) can decrease the contact angle of droplet[3-6]. The direct EW voltage is applied to the interface of the droplet and the electrode, the droplet contact angle is reduced, and only a few volts are usually required[7, 8]. However, direct EW usually uses a high electrolyte concentration, accompanied by intense electrolysis, when the voltage increases. EWOD covers the top of the electrode with a dielectric layer, which can effectively eliminate electrolysis[9]. The dielectric layer and hydrophobic topcoat of the EWOD are currently the subjects of extensive research, which increases the actuation voltage (approximately 100 volts) and potentially destabilizes due to breakdown[10]. Furthermore, electrodewetting uses ionic surfactants to apply only $\pm 2.5$ volts to a hydrophilic and conductive bare silicon surface to increase the contact angle by nearly 30°[2]. We realized that the initial contact angle of the droplets on the hydrophilic surface was greater than 30°, which should allow for further electrowetting. Suppose the bidirectional electric field is applied to the same conductive substrate, which can both achieve electrowetting and electrodewetting. In that case, the contact angle changes of droplets wetted and dewetted can be up to 50°, and more efficient droplet manipulation will be accomplished. Moreover, this mechanism can simplify the construction of digital microfluidic devices and facilitate the realization of equipment miniaturization, expanding the application range of digital microfluidic devices.

To study the ultimate principle and fundamental feature of the electrowetting and electrodewetting mechanism, we adopted a test method suitable for electrowetting, as shown in Fig. 1. An ionic surfactant is contained in the droplet, which includes a hydrophilic head and a hydrophobic tail, and distributed on the surface of the droplet. Fig. 1a, d shows the contact angle of the droplet (pH $\approx$ 7) containing dodecyltrimethylammonium bromide (DTAB) on a highly doped silicon wafer, chosen for its smooth surface without an external electric field, which this silicon wafer is specially processed, see the methods for details. When an electric field is applied between the droplet and the conductive substrate, ionic surfactant molecules migrate towards or away from the surface of the substrate near the contact line. A direct-current (d.c.) voltage is applied between the platinum wire inserted into the droplet and the substrate. When a forward electric field is applied between the platinum wire and the substrate (Fig. 1b, e), due to the use of DTAB as a cationic surfactant, surfactant molecules migrate toward the surface of the substrate. They are adsorbed on the surface of the substrate, mainly near the three-phase contact line, making the substrate hydrophobic and increasing the droplet contact angle. On the contrary, when a reverse electric field is applied between the platinum wire and the substrate (Fig.1c, f), cationic surfactants molecules migrate away from the surface of the substrate and tend to accumulate at the top of the droplet, and the anions in the water droplet form an electric double layer with the substrate, making the droplet further wet the substrate. Wetting and dewetting are reversible, and it is feasible to go from the forward electric field to the reverse electric field or vice versa.

Three kinds of ionic surfactants were used in the characterization experiment shown in Fig. 2, including two cationic surfactants and one anionic surfactant: DTAB, cetyltrimethylammonium bromide (CTAB), and sodium dodecyl sulphate (SDS), respectively. The highly doped silicon surface is negatively charged, and the aqueous droplet of ionic surfactant will appear to have an 'autophobing' effect[11-13], and the spontaneous adsorption of surfactant on the substrate surface increases the water droplet contact angle. Herein, the initial contact angle of a DTAB-containing aqueous droplet (about 5 $\mu l$, pH $\approx$ 7) exceeds 30°, at the isoelectric point however, the initial contact angle of an aqueous droplet (pH $\approx$ 2) is less than 25°. In order to eliminate this autophobing effect possibility, the aqueous droplet (pH $\approx$ 2) was used in the characterization experiments. As shown in Fig. 2a, the relationship between the contact angles of the three surfactants with the change of concentration using a voltage of $\pm 3$ V, the contact angle increases or decreases notably in the concentrations range which does not exceed the critical micelle concentration (CMC), especially at 0.05 CMC. For dewetting effect, we hypothesize that when ionic surfactant concentration is deficient, there are too few molecules to migrate toward the surface of the substrate. Thus the dewetting impact is not apparent, whereas when the surfactant concentration is close to or reached the critical micelle concentration, the aqueous droplet surface will cover a dense layer of surfactant molecular, there is a strong repulsive force between these molecules, molecules adsorbed to the substrate (mostly near the contact line) will rivet other molecules, making it difficult for the molecules on the droplet surface to migrate to the substrate, and making dewetting difficult. For electrowetting effect, it is mainly due to the electric double layer. The difference is that the resistivity (<0.005 $\Omega \cdot cm$) of highly doped silicon is much higher than most metals, and there is a thin enough oxide layer on the surface, which leads to the increase of capacitance[14]. Therefore, under such a low ion concentration, hydrolysis will not occur when the voltage is less than 3 V. Fig. 2b shows the relationship between contact angle variation and voltage action time. When the electric field is loaded with 0.3 s, the contact angle will change effectively, and the value can be more than 20°, which is slower than EWOD[15]. Fig. 2c shows the variation of contact angle with external electric field voltage. When the voltage is not more than 5 V, the change of contact angle increases with the voltage increase. When the voltage exceeds 5 V, obvious hydrolysis will occur.

To further investigate the advantages of this polarity-dependent electro-wetting or -dewetting, the aqueous droplet containing sodium bromide (NaBr) was used in the experiment. As shown in Fig. 3, two water droplet types are used. Fig. 3a shows an aqueous droplet with a concentration of 1.46 mM DTAB (0.1 CMC, pH $\approx$ 2) with an initial contact angle of 21°, Under the action of the applied electric field, electrowetting and electrodewetting can be realized, respectively. Fig. 3b shows the use of NaBr-containing droplet of the same concentration (pH $\approx$ 2) with an initial contact angle of 25°, which can only achieve electrowetting but not electrodewetting under the action of the external electric field. To further characterize the reversibility of the device prepared by this polarity-dependent electro-wetting or -dewetting mechanism, alternating current (a.c.) voltage was used to test these two aqueous droplets solution with DTAB and NaBr at 1.46 mM, as shown in Fig. 3c. The results show that the DTAB-containing droplet contact angle changes periodically with the electrical signal, which confirms its reversibility. The aqueous droplet containing NaBr (1.46 mM) does not change periodically, indicating that the change is irreversible. Whereas the contact angle changes of DTAB-containing droplet more significantly in the first period, then decreases and tends to be stable. This indicates that the electrodewetting is a dissipative process, but this does not affect the effective droplet drive, the change of droplet contact angle is still close to 30°.

To access the advantage of this technology, we have developed a digital microfluidic device (Fig. 4a). Using a water droplet contains 0.05 CMC DTAB In an open environment, without the use of any filler media and cover plates, We have compared droplet motion actuated by two mechanisms: the polarity-dependent electro-wetting or dewetting and electrodewetting, respectively. As shown in Fig. 4b, c, the proposed mechanism is verified. In addition, we characterized this droplet actuation mechanism and compared it with single electrowetting-on-dielectric or electrodewetting, as shown in Fig. 4c. For this polarity-dependent electro-wetting or -dewetting, we have got droplet actuation in front and back (Fig. 4f, h). For single electrowetting-on-dielectric or electrodewetting, the droplet actuation is only the front actuation or rear actuation (Fig. 4e, d, g). Therefore, the droplet moving mode with this polarity-dependent electro-wetting or -dewetting is more efficient, regardless of the absolute value of the contact angle change. Although the contact angle of the droplet will have a significant value in the forward direction under the action of a series of factors in the movement process. However, the droplet always moves in the hydrophilic direction, and the contact angle of the droplet at the forward end is usually smaller at the beginning of the droplet movement. Furthermore, the digital microfluidic devices fabricated by this polarity-dependent electro-wetting or -dewetting will have more stable droplet actuation and high working efficiency under the same contact angle variation.

## Methods

Microfluidics, which handle small amounts of fluids, are widely used in biomedicine and chemical synthesis[16-20]. Although high throughput can be achieved with microfluidic channels, miniaturization is challenging. Digital microfluidics using electrical signals to operate individual droplets can be performed precisely on miniaturized equipment[21]. Early electrowetting-on-conductor (EWOC) relied on the electric double layer to achieve wettability transformation[5, 22, 23]. In the last 20 years, EWOD has developed by leaps and bounds, mainly relying on the dielectric layer to change wettability, but the required voltage has also increased by nearly 100 times. The addition of a dielectric layer between the droplet and the electrode makes the digital microfluidic device come true. However, due to the rapid increase of voltage and the imperfection of thin film deposition on the electrode surface, this inevitably brings stability problems to the device. Recently, ionic-surfactant-mediated electrodewetting has been successfully implemented for digital microfluidics[2]. A technique combining both electrowetting and electrodewetting mechanisms has been implemented in this paper.

Contact angle measurement in this study was made using the sessile drop with the SL200KS tensiometer, produced in Shanghai SoLon. The contact angle is calculated by fitting the Yang-Laplace equation. The small contact angles (< 10°) can be measured using ImageJ with Brugnara's plugin[24]. The contact angle was measured using a procedure in which a 100 $\mu m$ diameter platinum wire (> 99.999%) was inserted into the middle of the aqueous droplet, about 100 $\mu m$ from the substrate, and the platinum wire should be cleaned with deionized water after each test. The substrate material is a 4-inch highly doped conductive silicon wafer (p-type, resistivity <0.005 $\Omega \cdot cm$). The substrate needs to go through the necessary cleaning steps before testing: piranha clean with 3 parts of 98% sulfuric acid and 1 part of 30% hydrogen peroxide at 120 °C for over 10 min; 50% hydrofluoric acid wash for 3 minutes; the wafer is rinsed with deionized water using Spin Rinse Dryer (SRD), then blown clean and heating drying. In this process, a natural oxide layer (<2 nm) has been formed on the silicon wafer's surface, laying a foundation for the later electrowetting. The whole test was carried out in a clean room, where the oxidation rate of silicon was very slow[25]. With the formation of the oxide layer, it was not easy to further oxidation in a conventional environment. As a result, the substrate can maintain surface properties in a clean environment for a long time, and function effectively for a long time. In tests, the cleaned silicon wafers remained effective in a clean room after two months.

A 5 $\mu l$ droplet was used for each test in the conventional laboratory environment, Fig. 2 shows the average of the three measurements. For the selection of ionic surfactants, we use two cationic surfactants: DTAB and CTAB, which have a bromine cationic head with one string of hydrocarbon tail. Therefore, an aqueous droplet containing NaBr was selected to compare with DTAB aqueous droplet to verify the electrowetting caused by the electric double layer, as shown in Fig. 3. We also set an anionic surfactant (SDS) to confirm the reliability of this phenomenon further. We dissolved these powders (Sigma Aldrich) of DTAB, CTAB and SDS in deionized water at room temperature, and their CMC is 14.6 mM, 0.92 mM and 8.2 mM, respectively[26]. Although individual contact angles may be different under the influence of the environment in the experiment, the increase and decrease of contact angles caused by the applied electric field are evident, which also reflects the robustness of the mechanism.

The demonstration digital microfluidics device shown in Fig, 4 was fabricated with a silicon-on-insulator (SOI) wafer of a 2-$\mu m$-thick top silicon layer (heavily-doped p-type, < 0.003 $\Omega \cdot cm$) and 2-µm-thick embedded silicon dioxide on an approximately 350-$\mu m$-thick base silicon wafer (lightly-doped n-type, 1–10 $\Omega \cdot cm$). First, the silicon electrode was defined with a 20-$\mu m$ gap between them by patterning the top silicon layer using reactive-ion etching with AZ5214 photoresist as the etching mask. The top layer silicon electrode was thinned down to about 800 nm by the reactive-ion etching after removing the photoresist, and then we packaged it on a printed circuit board. Each silicon electrode is independently addressed and voltage switched by LabView programming.

## Conclusions

We demonstrate ionic surfactant-mediated electrodewetting and electrowetting mechanism, which droplet dewetting or wetting on the surface of a conductive substrate in the air by changing the direction of the applied electric field. Considering the practicability of this mechanism, three critical parameters are tested in detail: the ionic surfactant concentration in the droplet, the voltage action time and the actuation voltage. The results show that this mechanism is reliable and robust. A further comparison was made between droplets of sodium bromide dissolved under the same conditions. On the one hand, it verifies the explanation of the mechanism; on the other hand, it also reflects the reversibility and controllability of the system under the action of these electrodewetting and electrowetting. Finally, considering the droplet actuation efficiency of the device, three droplet actuation mechanisms were compared respectively. This polarity-dependent electro-wetting or -dewetting mechanism has a more efficient droplet actuation mode, and the droplet actuation stability will be more guaranteed under the same conditions.

## Data availability

The data supporting this study are provided in the source data file. Source data are provided with this paper.

## Acknowledgments

We thank the Soft Matter Nanofab (SMN180827) of ShanghaiTech University.

## Author contributions

L.Z. and Y.L. conceived the ideas and designed the experiments. L.Z. performed experiments and analyzed the experimental data. L.Z., Y.L., and R.Z. wrote the manuscript. H.W. offered technical help.

## Competing interests

The authors declare no competing interests.

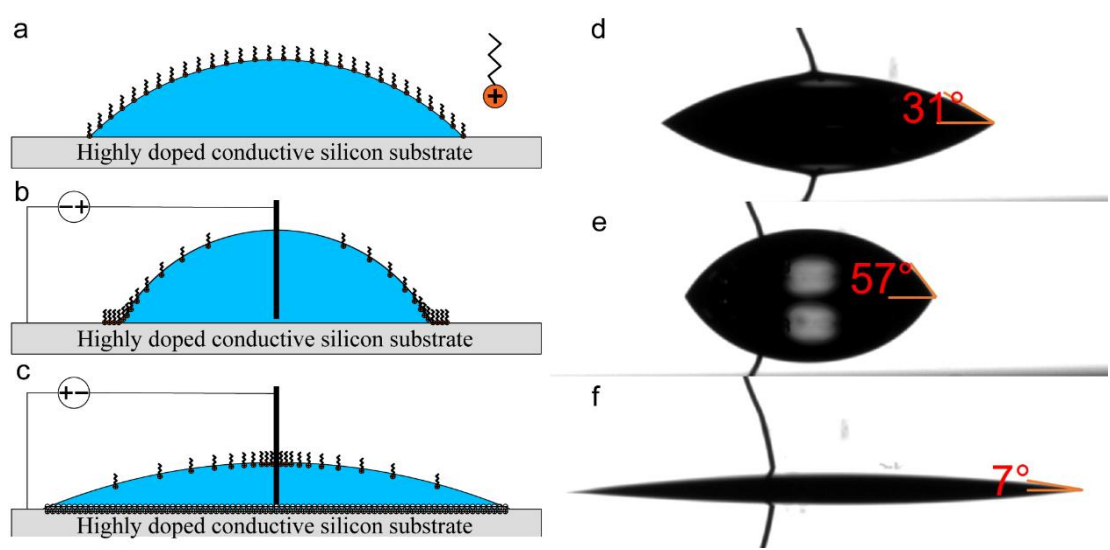

**Fig. 1 The proposed ionic-surfactant-mediated electrowetting and electrodewetting mechanism studied with a sessile drop on a conductive, hydrophilic substrate. a,** In the initial droplet state without an electric field, ionic surfactant molecules are mainly distributed at the liquid-gas interface. **b,** An electric field is applied to the liquid-solid interface, and the surfactant molecules (a cationic surfactant is shown) migrate to the substrate surface (mostly near the contact line), rendering the surface hydrophobic and making the droplet contact angle increase. **c,** An reverse electric field is applied to the liquid-solid interface, the ionic surfactant molecules migrate back to the liquid-gas interface, and the liquid-solid interface forms a double electric layer. The surface first returns to the initial hydrophilic state and then makes the droplet contact angle further decrease due to the double electric layer. **d, e, f,** the droplet experiment corresponds successively to a, b, and c, where an aqueous droplet containing DTAB (about 5 $\mu l$) is applied with a voltage of $\pm 5$ V on a highly doped silicon wafer.

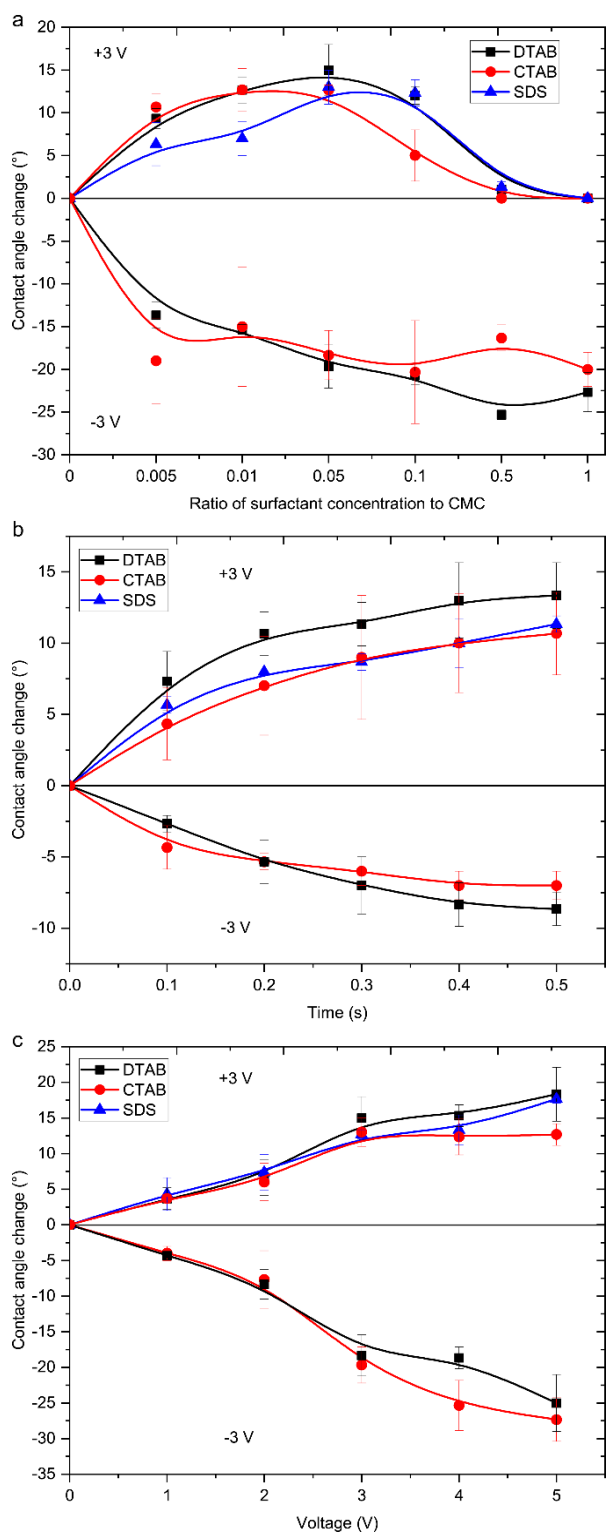

**Fig. 2 Effect of surfactant concentration, electric field action time and voltage on the change of droplet contact angle. a**, The droplet contact angle change with the concentration of cationic (DTAB, CTAB) and anionic (SDS) surfactants at the voltage of ±3 V. Due to the initial contact angle of the droplet containing SDS being too low (<10°), the contact angle reduction test caused by wetting was not carried out. **b**, Experiment using three surfactants at 0.05 CMC (effective concentration from Fig.2a) showed the relationship between the change of contact angle and the action time of electric field, the effective contact angle change can be achieved within 0.5 s. **c**, Three surfactants (0.05 CMC) were used in these tests

showing that the change values in contact angle were proportional to the increase in voltage. When the voltage exceeds 5 V, violent hydrolysis occurs.

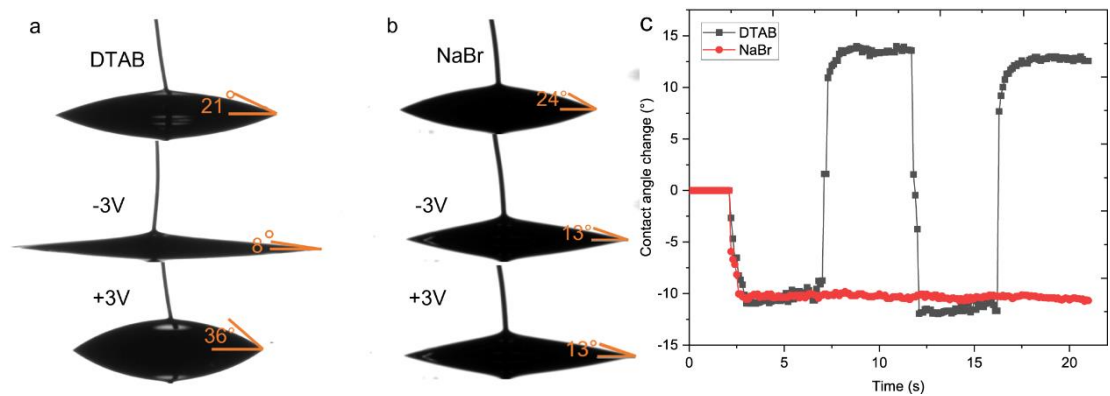

**Fig. 3 Ionic surfactants in this polarity-dependent electro-wetting or -dewetting mechanism can change the droplet contact angle reversibly and eliminate the memory effect. a,** Tests using droplets containing DTAB (1.46 mM) to wet and dewet the substrate under two opposing electric fields both at the voltage of 3 V. **b**, Experiments using droplets containing NaBr (1.46 mM) to wet the substrate, whereas this was irreversible and without electrodewetting. **c,** The droplet contact angle changes under the action of ac electrical signal, the a.c. electrical signal (5 V, 0.1 Hz) is applied between the droplet and the substrate. The aqueous droplet containing DTAB (1.46 mM) contact angle changes periodically, which confirms its reversibility. The aqueous droplet containing NaBr (1.46 mM) does not change regularly, indicating that the change is irreversible.

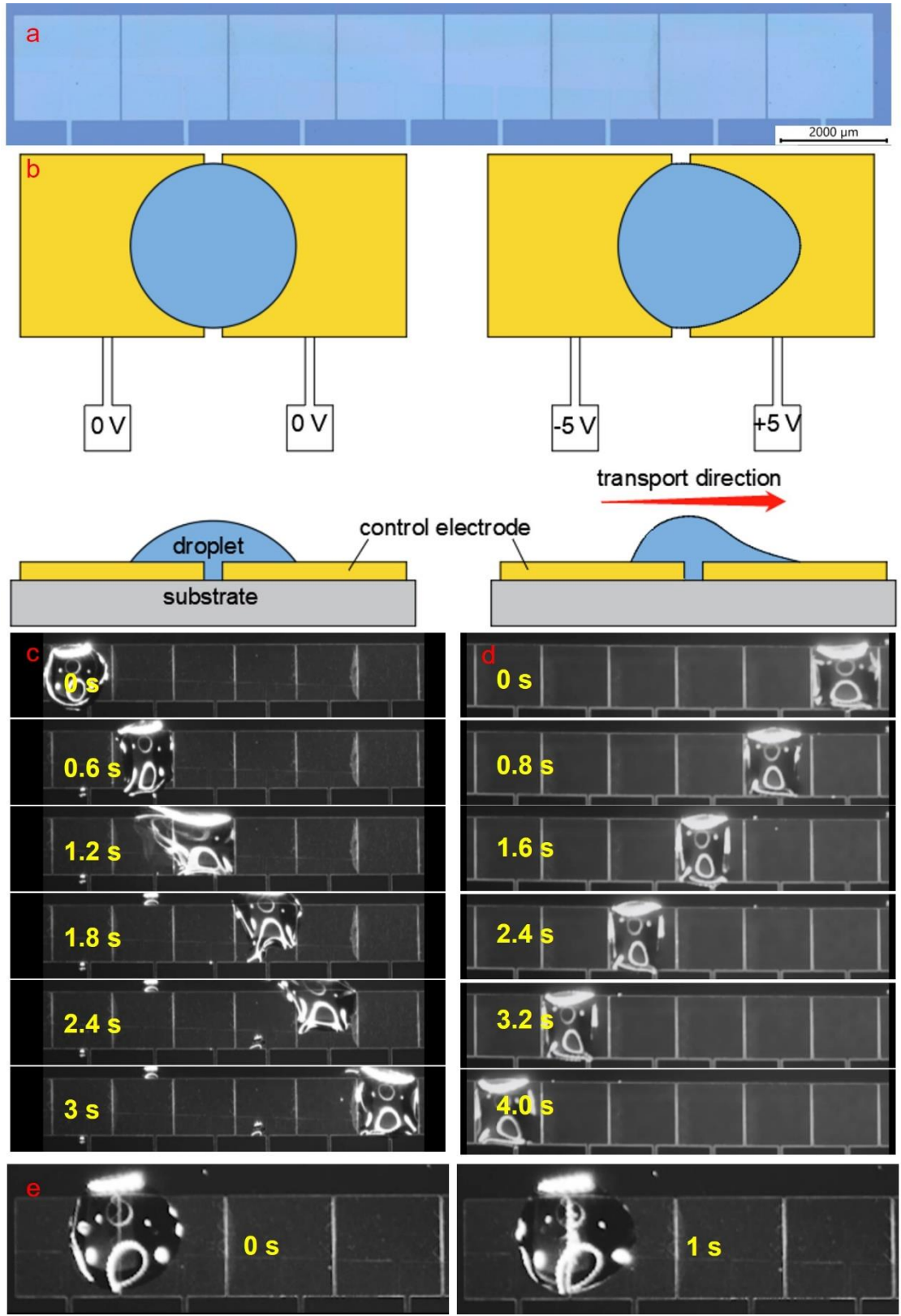

Fig. 4 Digital microfluidic device was constructed and three mechanisms of droplet transportation were compared, polarity-dependent electro-wetting or -dewetting, electrodewetting and electrowetting, respectively. The droplet transportation is realized by the polarity-dependent electro-wetting or -dewetting and electrodewetting. **a,** Light micrograph of the device. **b,** Cartoon schematic of the proposed mechanism, under the action of two poles voltage, electrowetting and electrodewetting co-occur, and one end of the droplet contact angle increases and the other end decreases, which actuation the droplet to move towards the hydrophilic end. **c, d,** Rate measurement of droplet transportation driven by the polarity-dependent electro-wetting or -dewetting and electrodewetting mechanism, the size of a single pixel is 2 $\mu m$ × 2 $\mu m$. The water droplet contains 0.05 CMC DTAB (about 0.8 $\mu l$) are used, and the droplet movement rate is about 3.33 $\mu m/s$ and 2.5 $\mu m/s$, respectively. **e,** The water droplet contains 1.46 mM NaBr is used, when a voltage is applied, the electrowetting occurs, but the droplet cannot be transported, causing the device to fail to work.